\documentclass{emulateapj}
\begin{document}

\newcommand{\tbol}{\mbox{$T_{bol}$}} 
\newcommand{\lbol}{\mbox{$L_{bol}$}} 
\newcommand{\lsmm}{\mbox{$L_{smm}$}} 
\newcommand{\msun}{\mbox{M$_\odot$}}
\newcommand{\mdisk}{\mbox{M$_{disk}$}}
\newcommand{\mstar}{\mbox{M$_*$}}

\title{The Diskionary: A Glossary of Terms Commonly Used for Disks
and Related Objects, First Edition}

\author{Neal Evans\altaffilmark{1} (NJE), 
Nuria Calvet\altaffilmark{2} (NC), 
Lucas Cieza\altaffilmark{3} (LC), 
Jan Forbrich\altaffilmark{4} (JF),
Lynne Hillenbrand\altaffilmark{5} (LH), 
Charlie Lada\altaffilmark{4} (CL), 
Bruno Mer{\'{\i}}n\altaffilmark{6} (BM), 
Steve Strom\altaffilmark{7} (SES), 
Dan Watson\altaffilmark{8} (DW)}

\altaffiltext{1}{Department of Astronomy, University of Texas at Austin,
  1 University Station C1400, Austin, TX~78712;nje@astro.as.utexas.edu}
\altaffiltext{2}{Department of Astronomy, University of Michigan, 830 
Dennison Building, 500 Church Street, Ann Arbor, MI 48109; ncalvet@umich.edu}
\altaffiltext{3}{Institute for Astronomy, University of Hawaii, Manoa, 
HI 96822.; Spitzer Fellow; lcieza@ifa.hawaii.edu}
\altaffiltext{4}{Harvard-Smithsonian Center for Astrophysics, 60 Garden 
Street, Cambridge, MA 02138; jforbrich@cfa.harvard.edu, clada@cfa.harvard.edu}
\altaffiltext{5}{Department of Astronomy and Astrophysics, California Institute of Technology, Pasadena, CA 91125; lah@astro.caltech.edu}
\altaffiltext{6}{Research and Scientific Support Department, European
Space Agency (ESTEC), PO Box 299, 2200 AG Noordwijk, The Netherlands;
bmerin@rssd.esa.int}
\altaffiltext{7}{National Optical Astronomy Observatories, 950 N. Cherry Ave., 
Tucson, AZ 85719; strom@noao.edu}
\altaffiltext{8}{Department of Physics and Astronomy, University of Rochester, 
Rochester, NY 14627-0171; dmw@astro.pas.rochester.edu}

\begin{abstract}

Based on our discussions for a panel discussion, we provide some definitions
of common usage of terms describing disks and related objects. 
\end{abstract}

\section{Introduction}

The authors were members of a panel leading a discussion at the meeting,
``New Light on Young Stars: Spitzer's View of Circumstellar Disks", which
was held in Pasadena, California in 2008, October 26-30. The motivation for 
the panel discussion was a proliferation of terms related to new discoveries
by Spitzer. The high sensitivity and spectral capabilities of Spitzer have
allowed many more disks to be found and studied in detail, revealing new
kinds of rare objects. Different groups have adopted a variety of names
for these new objects. The panel attempted to collect and clarify the 
meanings of these terms.

The contents of this document 
are based on our discussions in teleconferences and emails before the
meeting and on comments received from audience members during and after the
meeting. As our discussion proceeded prior to the meeting, we discovered 
that we often had very different conceptions of what various terms meant.
Rather than argue about who was correct, we instead tried to sharpen and
clarify the various definitions that were in use by panel members. Then
we added other terms that we knew to be used by others.

We want to make clear at the outset what this document is {\bf not}.
We do {\bf not} attempt to reject or exclude any common usages in favor 
of others;
we merely record them. In some cases, we {\bf do} comment on usages or names
that we found problematic, but we still include those as definitions.
Consequently, this clearly has the form of a dictionary, not a document
of ``permitted" and ``forbidden" usage. Since the focus was on disks,
we adopted the title of ``diskionary", which is easier to write than to
pronounce.

We take our model from Webster, as in

\noindent {\bf car}
\begin{enumerate}
\item a vehicle moving on wheels.
\item a vehicle adapted to the rails of a railroad or street railway
\item an automobile
\item (archaic) carriage, cart, chariot
\end{enumerate}

which illustrates that some definitions are more general and some are
specific subsets of the general definition.

In some cases, we include ``descriptions" which are not part of the definition
(i.e., necessary conditions), but which may describe many members of the
class. Some terms also have comments that capture some of our discussions.
Sometimes we have been able to supply references to the places where the
definition originated or was explained. We attempted to clarify whether
the term was observational, usually referring to some aspect of SED, etc.,
or physical, referring to an actual physical structure. However, many
of these definitions still mix these two categories.
The initials of the person who originally supplied the definition follow 
each term, but many iterations by various co-authors have occurred.
The definitions are not in alphabetical order, but instead we group
together related terms.
We begin with definitions that address SED classes, evolutionary stages,
and types of stars, as these may be used in the disk definitions.

\section{The Diskionary}

\noindent {\bf Class 0}
(NJE)
\begin{enumerate}
\item Observational, based on the SED properties, specifically $\tbol < 70$ K  
(Chen et al. 1995, ApJ, 445, 377) 
\item Observational, based on the SED properties, specifically $\lsmm > 0.005$
 \lbol.
(Andre et al. 1993)
\item (Original) Observational, a source that satisifies three criteria 
providing
indirect evidence of a central YSO in a centrally peaked submillimeter
core, with definition 2 also satisfied.  (Andre et al. 2000, PPIV, 59)
The conditions are: (a) Indirect evidence for a central YSO, as 
indicated by, e.g., the detection of a compact centimeter radio 
continuum source, a collimated CO outflow, or an internal heating source;
(b) Centrally peaked but extended submillimeter continuum emission
tracing the presence of a spheroidal circumstellar dust envelope
(as opposed to just a disk); (c) High ratio of submillimeter to bolometric 
luminosity, suggesting that the envelope mass exceeds the central stellar 
mass: $\lsmm/\lbol > 0.5$\%, where \lsmm\ is measured longward of 350 \micron. 
In practice, this often means an SED resembling a single-temperature 
blackbody at $T \sim 15-30$K.

\noindent Comment: The original definition emphasized {\bf indirect} evidence 
for a central
luminosity source, but most Class 0 sources can be seen in the mid-infrared
with Spitzer sensitivity, thereby providing {\bf direct} evidence and obviating
the need for some of the indirect evidence.
\end{enumerate}

\noindent {\bf Class I}
(NJE)
\begin{enumerate}
\item Observational, based on the SED properties, specifically 
  $\alpha \geq 0.3$, but $\tbol \geq 70$ K. ($\alpha$ is slope of $\nu S_{\nu}$ 
  between NIR and MIR, originally 2 and 20 \micron).
  (Lada 1987, IAU 115, 1, as modified by Chen et al. 1995)
\item Observational, based on the SED properties, specifically 
   $\alpha \geq 0.3$ with no constraint on \tbol. 
   (Greene et al. 1994, ApJ, 434, 614)
\item  Observational, based on the SED properties, specifically $\alpha > 0$
   (used when ``Flat" is omitted from the system) (Lada 1987)
\item Physical, an object whose emission is dominated by the envelope
   (Lada 1987) 
\end{enumerate}

\noindent {\bf Flat SED}
(NJE)
\begin{enumerate}
\item Observational, based on the SED properties, specifically 
  $-0.3 \leq \alpha < 0.3$. ($\alpha$ is slope of $\nu S_{\nu}$
  between NIR and MIR, originally 2 and 20 \micron).
  (Greene et al. 1994)
\item Observational, based on the SED properties, specifically 
   $350 < \tbol < 950$ K, or $500 < \tbol < 1450$ K if fluxes corrected for
   extinction (Evans et al. 2009, ApJS, in press)
\item Physical, an object whose emission is composite, arising from both 
  a disk, and an envelope; envelope likely possesses a wind-carved cavity.
\end{enumerate}

\noindent {\bf Class II}
(NJE)
\begin{enumerate}
\item Observational, based on the SED properties, specifically 
  $-1.6 \leq \alpha < -0.3$. ($\alpha$ is slope of $\nu S_{\nu}$ 
  between NIR and MIR, originally 2 and 20 \micron).
  (Greene et al. 1994)
\item Observational, based on the SED properties, specifically 
  $-2 \leq \alpha \leq 0$. ($\alpha$ is slope of $\nu S_{\nu}$ 
  between NIR and MIR, originally 2 and 20 \micron).
  (Lada 1987)
  (used when ``Flat" is omitted from system)
\item Physical, an object whose emission is dominated by the star and disk
   (Lada 1987)
\item  Physical, a star surrounded by an accretion disk 
\end{enumerate}

\noindent {\bf Class III}
(NJE)
\begin{enumerate}
\item Observational, based on the SED properties, specifically 
  $\alpha < -1.6$.  ($\alpha$ is slope of $\nu S_{\nu}$ 
  between NIR and MIR, originally 2 and 20 \micron).
   (Green et al. 1994)
\item Observational, based on the SED properties, specifically 
  $\alpha \leq -2$.  ($\alpha$ is slope of $\nu S_{\nu}$ 
  between NIR and MIR, originally 2 and 20 \micron).
  (Lada 1987)
\item Physical, an object whose emission is dominated by the star.
   (Lada 1987)

Description: A pre-main sequence star exhibiting no evidence
for a primordial/accretion disk but which may have a small fractional
infrared excess arising from a remnant primordial/quiescent or
secondary/debris disk.
\end{enumerate}

\noindent {\bf Stage 0}
(NJE)
\begin{enumerate}
\item Physical, mass in envelope exceeds mass in disk plus star
   (based on physical version of Class 0 definition by Andre et al.)
\end{enumerate}
\noindent Description: Class 0 sources are often, but not always, Stage 0 sources.

\noindent {\bf Stage I}
(NJE)
\begin{enumerate}
\item Physical, mass infall rate over mass in envelope at least $10^{-6}$ 
  per year (includes Stage 0) (Robitaille et al. 2006, ApJS, 167, 256)
\item Physical, mass in envelope less than mass in star plus disk, but
  greater than 0.1 \msun. (Crapsi et al. 2008, Astr. Ap., 486, 245)
\end{enumerate}
\noindent Description: Class I sources are often, but not always, Stage I sources.

\noindent Comments: These definitions, whether by mass or by relative infall 
rate may have problems in describing more massive stars.

\noindent {\bf Stage II}
(NJE)
\begin{enumerate}
\item Physical, A young star or protostar with less than 0.1
        \msun\ remaining in an envelope, but with a primordial disk
\item Physical, mass infall rate over mass in envelope $< 10^{-6}$ per year
    and $\mdisk/\mstar > 10^{-6}$ (Robitaille et al. 2006, ApJS, 167, 256)
\end{enumerate}
\noindent Description: Class II sources are often, but not always, Stage II sources.

\noindent Comments: Again, there may be issues for more massive stars.
Perhaps one could generalize by requiring an envelope mass $\leq 0.1 
  (\mstar + \mdisk)$.

\noindent {\bf Stage III}
(NJE)
\begin{enumerate}
\item Physical, a young star without a primordial disk
\item Physical, mass infall rate over mass in envelope $< 10^{-6}$ per year
    and $\mdisk/\mstar < 10^{-6}$ (Robitaille et al. 2006, ApJS, 167, 256)
\end{enumerate}
\noindent Description: Class III sources are often, but not always, Stage III sources.

\noindent Comments: Again, there may be issues for more massive stars. The committee
is divided on whether to say ``accretion" rather than ``primordial" in 
definition \#1.

\noindent {\bf Group I}
(LH)
\begin{enumerate}
\item Observational definition, albeit interpreted via theoretical models,
pertaining to Herbig Ae/Be stars with SEDs that are loosely consistent with
geometrically flat, optically thick accretion disks but having small inner 
gaps or holes (Hillenbrand et al. 1992, ApJ, 397, 613).
\end{enumerate}
\noindent Description: These objects are analogous to the Class II object
definition typically applied to lower mass CTTS, but the definition here is 
much less quantitative. The inner hole may correspond to the dust destruction
radius.

\noindent {\bf Group II}
(LH)
\begin{enumerate}
\item Observational definition, albeit interpreted via theoretical models,
pertaining to Herbig Ae/Be stars with SEDs that exhibit mid-infrared emission
above that explainable with a geometrically flat optically thick accretion 
disk (Hillenbrand et al. 1992).
\end{enumerate}
\noindent Description: The excess emission was attributed to an envelope.
These objects are only vaguely similar to the Class I category as 
applied to lower mass CTTS. While Group I Herbig Ae/Be stars
and Class II CTTS are similar in their physical interpretation, Group
II Herbig Ae/Be stars are not necessarily analogous to Class I CTTS
in a quantitative sense.

\noindent {\bf Group III}
(LH)
\begin{enumerate}
\item Observational definition,  pertaining to Herbig Ae/Be stars with small
near-infrared to mid-infrared excesses that are far weaker than those 
associated with
geometrically flat, optically thick accretion disks and roughly consistent 
with those associated with Classical Be stars (Hillenbrand et al. 1992).
\end{enumerate}

\noindent {\bf group I}
(LH)
\begin{enumerate}
\item Observational, strong and almost flat or double-peaked 2-100 \micron\ SEDs
in ISO spectrophotometry of Herbig Ae/Be stars (Meeus et al. 2001, Astr. Ap., 
365, 476).  
Sub-categories Ia and Ib are defined by structure in
the 10 \micron\ silicate feature with Ib having essentially no broad 10 \micron\
emission bump.  Both Ia and Ib can exhibit narrow PAH emission, however.
\end{enumerate}
\noindent Description: These objects are roughly consistent with flared disk
models (Dullemond).

\noindent {\bf group II}
(LH)
\begin{enumerate}
\item Observational, weak and declining 2-100 \micron\ SEDs in ISO spectrophotometry
of Herbig Ae/Be stars (Meeus et al. 2001).  Often have 10 \micron\ silicate
emission.  Also can exhibit PAH emission, but not as strongly as in the group I
objects.
\end{enumerate}
\noindent Description:  These objects are roughly consistent with self-shadowed
disk models (Dullemond).

\noindent {\bf Classical T Tauri Star (CTTS)}
(SES)
\begin{enumerate}
\item (Original) Observational, a solar-type PMS star that exhibits 
at least one of the following:
(a) broad, strong $H\alpha$ emission (Joy, 1945, ApJ 102, 168); 
(b) ultraviolet excess emission (above photospheric levels); 
(c) forbidden and in some cases permitted line emission; and 
(d) irregular optical variability.

Description: Such stars can exhibit IR SEDs that span Class I-II,
and in some cases, Transition Disk SEDs (see below).
Typical examples are AA Tau; BP Tau (Class II); and HL Tau, DG Tau.
\item Physical, A solar-type PMS star surrounded by an accretion disk, and
in some cases, an accretion disk and a `remnant' envelope. 
An accretion-driven wind, manifest in emission line spectra, is also present.
\end{enumerate}

\noindent Description: The key distinguishing characteristics are the presence of an 
accretion disk and an accretion-driven wind. However, CTTS cover a 
`broad' range of evolutionary stages.

\noindent {\bf Herbig Ae/Be Stars}
(NC)
\begin{enumerate}
\item Observational, a category of spectral type A and B stars that exhibit properties similar
to those of CTTS: strong optical emission lines and infrared to millimeter
excesses consistent with disk emission.
\item (Orig.) Observational, Herbig (1962) first drew attention to a set of early type
objects with emission lines, associated with localized nebulosity and
larger scale dark clouds.  As is the case for CTTS/WTTS, the original
classification implied nothing about the presence of disks, though disks
were later recognized as the physical driver for many of the observational
characteristics.
\end{enumerate}

\noindent Description: These systems are interpreted as pre-main sequence and
zero-age main sequence objects, with ages comparable to CTTS and WTTS
but larger masses, $2 < \mstar/\msun < 10$.  They are often referred to 
as just ``Herbig Stars".

\noindent {\bf Weak-line T Tauri Star (WTTS)}
(SES)
\begin{enumerate}
\item (Original) Observational, a ``solar-type" ($M_* \sim 0.5 \msun$, K7-M0), 
age $\sim 1$ Myr, pre-main sequence star 
(from HRD location) that has $W(H\alpha) < 10$ \AA. 

Description: The limit was chosen in order to separate objects
which showed the signatures of optically thick accretion
disks (based on IRAS + NIR SEDs) from those that appeared
to be photospheric.

\item (More modern) Observational, any pre-main sequence star that 
shows no evidence of accretion based on an age and spectral-type
appropriate $W(H\alpha)$. 

Description: Firm inclusion in the class would
require additional information from high resolution, high
S/N spectra to test for the presence (CTTS) or absence
(WTTS) of $H\alpha$ profiles indicative of accretion, or
signatures (e.g., forbidden [OI] or [SII]) of winds driven by accretion.
\item Physical, a young star that is no longer surrounded by an
accretion disk.
\end{enumerate}

\noindent Comments: The definition of WTTS was introduced to link signatures of
accretion ($H\alpha$ emission; forbidden line emission) diagnosed from the
presence/kinematics of gas, with signatures of optically
thick disks provided by IR SEDs. The definition that was
applied to the initial Taurus sample is obsolete, in the
sense that the $W(H\alpha) = 10$ \AA\ limit is an appropriate (and only
approximate) discriminant for Taurus-age stars with masses $\sim 0.5$ \msun.
However, the idea of using both gas and dust diagnostics
to assess the physical state of the disk is still enormously valuable.
In particular, it is by using both SEDs and gas
diagnostics that we can probe the evolutionary state of
transition/anemic disks and assess/constrain the likely
physical cause for the observed SED signature.

\noindent {\bf Post-T Tauri star (PTTS)}
(SES)
\begin{enumerate}
\item Observational, a young star that exhibits a photospheric
spectrum over a wavelength range from the NIR and longward.

Description: Among solar-type PMS stars, such objects can be
identified from their location in the HRD, from their
x-ray properties supplemented by HRD location (confirmed
from parallaxes; radial velocities; clear membership in a
cluster.....) plus a robust SED. Among higher mass stars,
such objects might be called ``Post Ae/Be stars" and can be
identified via their location in the HRD in a
cluster/association plus robust SEDs.
\item Physical, a young star of any mass that no longer is
surrounded by a circumstellar accretion disk.
\item Physical, an evolutionary step between T Tauri Stars and ZAMS, 
generally thought to refer to stars on the radiative track.

\noindent Comments: In these definitions, there is no physical difference 
between a WTTS and a post-TTS. It's just the choice of observational `tool' for 
certifying that such an object is no longer accreting material.
There is some sentiment on the committee for specifying that
a Post-T Tauri star is older than either a CTTS or a WTTS (e.g., with
ages of 5-30 Myr). See also naked T Tauri star.
\end{enumerate}

\noindent {\bf Naked T Tauri star (NTTS)}
(supplied by Scott Wolk and Fred Walter)
\begin{enumerate}
\item Observational, a low mass pre-main sequence star that exhibits no 
evidence of an optically thick circumstellar disk or on-going accretion.

\noindent Description: In this picture, a NTTS differs from a post-T Tauri star 
in being an observational definition while the PTTS is described by 
definition 3 for post-T Tauri Star.
\end{enumerate}

\noindent {\bf G-type T Tauri Star (GTTS)}
(LH)
\begin{enumerate}
\item Observational,  G and K0 type stars that are similar to T Tauri stars

\noindent Description: Used to distinguish them from lower mass K-M traditional 
TTS and the higher mass Herbig Ae/Be stars. (Herbst and Shevchenko,
1999, AJ, 118, 1043)
\end{enumerate}

\noindent {\bf Intermediate-Mass T Tauri Star (IMTTS)}
(NC)
\begin{enumerate}
\item Observational,  T Tauri stars, with masses between 1.5
\msun\ and 4 \msun\ 

\noindent Description: The predecessors of Herbig Ae/Be stars
(Calvet et al. 2004, AJ, 128, 1294)
\end{enumerate}

\noindent {\bf Young Stellar Object (YSO)}
(NJE)
\begin{enumerate}
\item Physical, an object in any of the Stages defined above.
   (Strom et al. 1975, ARAA, 13, 187)
\item Observational, an object in any of the Classes but with an infrared 
   excess detectable in Spitzer surveys (a definition used by the c2d team 
   to distinguish YSOs from the larger class of pre-main-sequence objects (PMS).
\end{enumerate}

\noindent {\bf Pre-Main-Sequence Object (PMS)}
(NJE)
\begin{enumerate}
\item Observational, a star that lies above the main sequence.
\end{enumerate}

\noindent Comment: This object may or may not have a detectable infrared excess. It
is identified by various indicators of youth. It includes CTTS, WTTS, PTTS,
GTTS, and NTTS.

\noindent {\bf protostar}
(NJE)
\begin{enumerate}
\item Physical, an object that is still accreting mass and that will become or
    already is a main sequence star.
\item Physical, an object deriving most of its luminosity from accretion and
   that will become a main sequence star.
\item Physical, the entire structure of central star, disk, and envelope.

\noindent Comments: The first definition allows the inclusion of massive 
protostars, 
which can be already burning nuclear fuel but still growing in mass.
The second does not allow them above a certain mass where the nuclear-burning 
luminosity exceeds that from accretion. Both the first two apply to the 
central object, while the third refers to the entire structure. For the
first two definitions, objects in Stages 0, I, or II could {\bf host} protostars;
For the third definition, only Stages 0 and I would {\bf be} protostars.
\end{enumerate}

\noindent {\bf protostellar disk}
(CL)
\begin{enumerate}
\item Physical, a primordial disk during the protostellar phase.
\item Physical, an accretion disk associated with a protostar and
through which the protostar (Def. \#1 or \#2) gains most of its mass.
\end{enumerate}

\noindent {\bf accretion disk}
(SES)
\begin{enumerate}
\item Observational, signified via at least one of the following:
(a) broad $H\alpha$ emission, which at appropriate resolution and S/N 
exhibits kinematic signatures of accretion (inverse P-Cygni);  or
(b) ultraviolet excess emission (above photospheric levels),
(c) gas in Keplerian rotation around the star; or
(d) large amplitude and irregular photometric variability.

Description: the excess UV is thought to arise from hot spots from
accreting matter. It is often combined with emission lines (e.g. 
CO, [NeII], HCN, ...) that exhibit the characteristic double-horned profiles 
indicative of gas that could supply the accretion. 
\item Physical, a disk -- isolated or fed by material infalling from a
protostellar envelope -- which is transporting material inward toward 
the stellar surface. 
\end{enumerate}

\noindent {\bf circumstellar disk}
(LH)
\begin{enumerate}
\item Physical, any disk of gas and/or dust surrounding a star.
\item Physical, a disk surrounding only a single star, not a binary.
\end{enumerate}

\noindent Comment: Definition \#1 includes all of the disks described here.
Definition \#2 excludes circumbinary disks.

\noindent {\bf primordial disk}
(LH)
\begin{enumerate}
\item Physical, gas-rich circumstellar disk composed of matter originating in the
interstellar medium but processed through the star/disk-forming accretion 
shock.
\item Physical, dust and gas that are present in a ratio close to 1:100 and 
relatively unprocessed despite the fact that the material is actively 
evolving through grain growth, settling, and agglomeration and 
chemical effects. 
Dust dynamics are controlled by coupling with the gas.
\item Physical, A young disk, possibly accreting and generally optically thick,
that is capable of forming planets now or in the future, i.e. protoplanetary.
\item (Orig.) The chemically unprocessed part of the Kuiper Belt beyond 50 AU.
\end{enumerate}

\noindent {\bf pre-transitional disk}
(DW)
\begin{enumerate}
\item Observational, a YSO with an infrared excess similar to the median
of its home cluster over most of the spectrum, but with a substantial
deficit compared to the median over some intermediate range of wavelengths.

\noindent Comments: ``home cluster" is  bit vague, and what about non-clustered YSOs?

\item Physical, a disk around a T Tauri star with an optically-thin gap
separating optically-thick inner and outer disk components that resemble,
in their excess flux, gap-less disks.

\noindent Comments: This group of objects exhibit specific SED characteristics
that are not covered by other more general terms.  However, the panel
is uncomfortable with the name, which implies an evolutionary sequence
that is not necessarily associated with this observational or physical
entity. Under the broad definition of a transitional disk (\#3), this is
just one form of a transitional disk. An alternative term has been suggested
for the physical definition: ``a disk with an annular gap".
\end{enumerate}

\noindent {\bf transitional disk}
(NC)
\begin{enumerate}
\item Observational, a disk with a deficit of flux relative to the median of 
Taurus in the near-IR and fluxes comparable or higher than the median at 
mid-IR and longer wavelengths.  

Description: These observational characteristics have been
interpreted in terms of an inwardly truncated disks with little small dust left
in the hole. The truncation has been attributed to several mechanisms,
including photoevaporation and formation of planets. 
(see K. Strom et al. 1989, AJ, 97, 1451 for first use of this term)

Comments: Further detailed studies of these disks have shown the
following variants: 
(a) disks with NO measurable dust or gas inside a hole of size $>>$ dust
destruction radius;
(b) disks with measurable, but optically thin dust within an inner hole
and extending inward to the dust destruction radius; 
(c) disks with measurable gas in Keplerian rotation extending inward 
from the inner hole; dust to gas ratios $<<$ 100.

\item Observational, a disk with fluxes lower than primordial disks
(represented by the median in Taurus) at all wavelengths with an excess.

\item Physical, a disk that satisfies the description of EITHER definition 
  1 or 2; that is, any disk in transition between primordial and debris disks. 

\end{enumerate}
\noindent Description:  These disks show significant effects arising from
(a) evolution of dust to larger sizes (this version might be an anemic disk); 
(b) the combined effects of accretion and photoevaporation; and 
(c) the dynamical effects of forming planets on the structure, 
  accretion rate and dust content of the inner disks 
  (b and c could be ``cold disks" or transitional disks in the more narrow
   sense of definition \#1)

\noindent Comments: This is one of the more confused terms. If definition \#3 is used,
 for transitional disk, it includes pre-transitional disk, transitional disks 
 in the narrow sense of definition \#1 (also cold disks), and anemic disks. 
 If you use it, be sure to define it.

\noindent {\bf annular disk}
(NJE)
\begin{enumerate}
\item Physical, a disk with an inner hole with a much reduced abundance of 
  dust.

Comments: This is a possible replacement term for definition \#1 of
transitional disks if the broader definition of transitional disk is used.
(If you asked a person on the street to name this shape, they would say 
``halo", but the cosmologists have already misused that term...)
\end{enumerate}

\noindent {\bf anemic disk}
(CL)
\begin{enumerate}
\item Observational, a disk whose observed flux at all wavelengths
is less than that of an optically thick, spatially flat accretion
or primordial (CTTS) disk
\item Physical, a homologously depleted primordial (CTTS) disk 
\item Physical, an evolved primordial (CTTS) disk
\item Observational, a disk with an IRAC slope intermediate between
that of a primordial CTTS disk and a debris disk. Can exhibit active
or passive accretion signatures.
\end{enumerate}

\noindent Comments: If definition \#3 is used for transitional disk, this is one
variant of a transitional disk.

\noindent {\bf debris disk}
(LH)
\begin{enumerate}
\item Physical, circumstellar disk composed of rock, dust, and ice.
\item Physical, gas-poor circumstellar disk in which the primary physical 
process is the collisional grinding down of planetesimals into smaller 
particles, eventually dust which is known as ``second generation" dust.
\item Physical, optically thin disk in which material is removed from the 
system via various drag or blow-out scenarios on timescales much 
less than the age of the star.
\end{enumerate}

\noindent Description:  Indicator of a planetary system given that large, 
planetary-mass bodies are required in order to induce and maintain the 
collisional cascade.  In our Solar System, the Asteroid Belt and the Kuiper 
Belt are well-separated debris belts that would be considered a 
tenuous debris disk if viewed from afar.

\noindent {\bf protoplanetary disk}
(CL)
\begin{enumerate}
\item Physical, a primordial disk with the potential to form planets.
\end{enumerate}

\noindent {\bf circumbinary disk}
(DW)
\begin{enumerate}
\item Physical, a binary system and a disk of extent significantly 
greater than the separation of the binary components.
\item Observational, an infrared excess in a binary system, for which SED or
images indicate an extent greater than the separation of the components
of the binary. 

\noindent Description: The two components may be various pairwise combinations of
stars, brown dwarfs, or planets. 
Each may have its own circumprimary or circumsecondary
disk, also lying within the circumbinary disk. The components may
truncate the circumbinary disk from within, in which case the system would
appear also as a transitional (Def. \#1) or cold disk.
\end{enumerate}

\noindent {\bf gaseous disk}
(BM)
\begin{enumerate}
\item Physical, a disk with a gas to dust mass ratio close to that in the ISM 
(100:1), typically associated with primordial T Tauri or Herbig Ae/Be disks.
\item Observational, a disk that shows evidence of gas to dust $>> 100$  via 
(a) observation of gas tracers such as CO, [Ne II]; or accretion signatures 
(broad $H\alpha$; inverse P-Cygni emission profiles; [O I], [S II], [N II], 
[Fe II] and similar wind signatures); and 
(b) no evidence of dust emission arising from heated small (0.1 to 10
\micron) grains.  
\item Observational, an object with small IR excess emission ($<<$ that expected 
from an optically thick disk), and with SEDs that mimic those observed 
around classical Be stars (which are thought to be surrounded by gaseous 
EX-cretion disks). 
\end{enumerate}

\noindent Description: Hillenbrand et al. (1992, ApJ, 397, 613) dubbed intermediate 
mass objects falling into this category ``Group III" Ae/Be stars. 

\noindent Comments: Note that these definitions are all quite different from one
another.  If you use this term, be sure to define it.

\noindent {\bf optically thick disk}
(BM)
\begin{enumerate}
\item Physical, a disk with an optical depth to its own radiation 
larger than one at most radii.

Description: Usually associated with an SED slope close to $-1$ in 
$\nu S_{\nu}$, or a substantial ratio of disk luminosity to stellar luminosity.
\item Physical, a disk that is optically thick at most radii to the 
radiation from the star.

Description: usually associated with an observation of $L_{disk}/L_{star}$ 
at least 0.1.
\end{enumerate}

\noindent Comments: Note that these two definitions are completely different. If you
use this term, be sure to define it.
Note that definition \#1 should specify the wavelengths being discussed or
should specify that one is referring to something like the Rosseland mean
opacity.

\noindent {\bf optically thin disk}
(BM)
\begin{enumerate}
\item Physical, a disk with an optical depth to its own radiation much smaller 
than 1 at most radii.
Description: It is usually associated with an SED slope close to $-3$ 
in $\nu S_{\nu}$, or a low ratio of disk luminosity to stellar luminosity.
\item Physical, a disk optically thin at most radii to the radiation from the 
   star.

\noindent Description: usually associated with an observation of 
$L_{disk}/L_{star} < 0.01$.

\noindent Comments: Note that these two definitions are completely different. 
If you use this term, be sure to define it. 
Note that definition \#1 should specify the wavelengths being discussed or
should specify that one is referring to something like the Rosseland mean
opacity.
\end{enumerate}

\noindent {\bf cold disk}
(LC)
\begin{enumerate}
\item Observational, based on the mid-IR flux ratio, specifically the
flux density ratio between 30 and 13 \micron: $S_\nu(30)/S_\nu(13) > 5$ 
(Brown et al. 2007, ApJL, 664, 107).

Description: These observational characteristics have been interpreted
in terms of disks with inner holes or wide gaps with little dust
($M_{dust} < M_{Moon}$) left in the hole/gap.
\item Observational, a primordial disk with little or no near-IR excess.
\item Observational, a disk whose SED rises steeply at wavelengths longer
than expected if the disk extended to the dust sublimation radius.
\item Physical, a truncated disk missing the hot inner disk.
\end{enumerate}

\noindent Comments: If the broad definition of transitional disk is used (Def. \#3), 
this is one form of a transitional disk, and the term cold distinguishes
it from anemic disks, for example. If definition \#1 for transitional disk 
is used, this is an observational synonym for transitional disk.
Some people did not like the name because many of the
``cold" disks are accreting and thus have some very hot gas. Others did not
like the term because ``cold" is a physical term, while a ``cold disk" 
is an observational definition. Here ``cold" refers to the SED.

\noindent {\bf hot/warm disk}
(LC)
\begin{enumerate}
\item Physical, relative term to indicate the region of the disk where
the temperature is higher/intermediate.
\item Physical, the innermost part of the disk/the warm surface layer
of a flared disk.
\item Observational, a disk whose SED shows near-IR excess.

\noindent Comments: The terms are very vague and not commonly used.
\end{enumerate}

\noindent {\bf flat disk}
(NC)
\begin{enumerate}
\item Physical, disk height is constant with radius
\end{enumerate}

\noindent Description: their spectral energy distribution is 
$\lambda F_\lambda \propto \lambda^{-4/3}$

\noindent {\bf flared disk}
(NC)
\begin{enumerate}
\item Physical, the ratio of disk height to radius increases with radius
\end{enumerate}

\noindent Description: their spectral energy distribution
  $\lambda F_\lambda$ is flatter (less steep) than $\lambda^{-4/3}$
  Often the dependence is more like $\lambda^{b}$ with $b$ between 0 and 1, but
 there are likely intermediate cases.

{\bf Parts of Disks:}

\noindent {\bf inner disk}
(DW)
\begin{enumerate}
\item Physical, the portion of a YSO disk lying closer to the central star
than the outer disk. 

\noindent Comments: This is a relative term, and the disk domain to
which it refers needs to be defined and limited to its context.
\end{enumerate}

\noindent {\bf outer disk}
(DW)
\begin{enumerate}
\item Physical, the portion of a YSO disk lying further from the central
star than the inner disk. 

\noindent Comments: This is a relative term, and the disk domain
to which it refers needs to be be defined and limited to its context.
\end{enumerate}

\acknowledgments
We thank the audience at the panel discussion for lively comments and
corrections. In particular, we thank L. Allen for finding some errors
and S. Wolk and F. Walter for supplying an additional definition.

\end{document}